\documentclass[aip,apl,showpacs]{revtex4-1}
\usepackage{amssymb}
\usepackage{setspace}
\usepackage{amsmath}
\usepackage{latexsym}
\usepackage{graphicx}
\usepackage{graphics}

\begin{document}

\title{Lyotropic chromonic liquid crystal semiconductors for
water-solution processable organic electronics.}
\author{V. G. Nazarenko}
\affiliation{Institute of Physics, prospect Nauky 46, Kiev-39, 03039, Ukraine}
\author{O. P. Boiko}
\author{M. I. Anisimov}
\author{A. K. Kadashchuk}
\affiliation{Institute of Physics, prospect Nauky 46, Kiev-39, 03039, Ukraine}
\author{Yu. A. Nastishin}
\affiliation{Liquid Crystal Institute and Chemical Physics Interdisciplinary
Program, Kent State University, Kent, OH 44242}
\affiliation{Institute of Physical Optics, 23 Dragomanov str., Lviv, 79005, Ukraine}
\author{A. B. Golovin}
\affiliation{Liquid Crystal Institute and Chemical Physics Interdisciplinary
Program, Kent State University, Kent, OH 44242}
\author{O. D. Lavrentovich}
\email{olavrent@kent.edu}
\affiliation{Liquid Crystal Institute and Chemical Physics Interdisciplinary
Program, Kent State University, Kent, OH 44242}
\date{\today }

\begin{abstract}
We propose lyotropic chromonic liquid crystals (LCLCs) as a distinct class of materials for organic electronics. In water, the chromonic molecules stack on top of each other into elongated aggregates that form orientationally ordered phases. The aligned aggregated structure is preserved when the material is deposited onto a substrate and dried. The dried LCLC films show a strongly anisotropic electric conductivity of semiconductor type.  The field-effect carrier mobility measured along the molecular aggregates in un-optimized films of LCLC V20 is $0.03$ $cm^{2}V^{-1}s^{-1}$.  Easy processibility, low cost and high mobility demonstrate the potential of LCLCs for microelectronic applications.
\end{abstract}

\maketitle

\bigskip \newpage

The field of organic electronics in general and organic field-effect transistors (OFETs) in particular has gained considerable attention because of the emerging applications in flexible electronics and displays \cite{Pisula,Grozema,Demenev,Adam}. As compared to silicon semiconductors, advantages of organic materials are solution-based processing, broader variety of device designs and often lower cost of fabrication. Especially promising are thermotropic liquid crystals (LCs) with their long-range orientational order, dense packing and adaptive structure that heals local defects \cite{Percec,Oikawa,Funahashi}. One first creates a LC precursor with a well aligned structure and then fixes it, typically by thermal quench and crystallization \cite{Garnier,Amundson}. There are two limitations \cite{Demenev}: (a) alignment of large-area monodomains is challenging and (b) the properties of thermotropic LCs are strongly temperature dependent. We propose to use the so-called lyotropic chromonic liquid crystals (LCLCs) \cite{Lydon,Tam-Chang,Collings} for OFET applications.

The polyaromatic cores of the LCLC molecules tend to stack face-to-face with a small separation of $\sim0.34$ nm, thanks to the strong $\pi-\pi$ attractions, forming elongated columnar aggregates \cite{Lydon,Tam-Chang,Collings}. Polar groups at the periphery of the cores make the aggregates soluble in water. At sufficiently high concentrations, the columns form a nematic. The orientational order with dense packing is preserved when the aggregates are transferred from the nematic water solutions into thin films by shear deposition and drying \cite{Schneider,Schneider_Artyushkova,Kaznatcheev,Nastishin,Ignatov}. Titled alignment of LCLCs is hindered by the aggregate structure \cite{Nazarenko}, thus the proper planar alignment can be achieved without special aligning layers. We show that dry aligned films of LCLC exhibit semiconducting properties with anisotropic mobility of charge carriers that is maximum along the aggregates.
\begin{figure}[b]
\caption{(Color online) Molecular structure of V20 (a). V20 material is a
mixture of isomers that differ in the location of sulfonate groups. OFET (b)
fabricated on SiO$_{2}/n$-Si substrates. }
\label{Molecular_structure}\centering
\includegraphics[width=0.48
\textwidth]{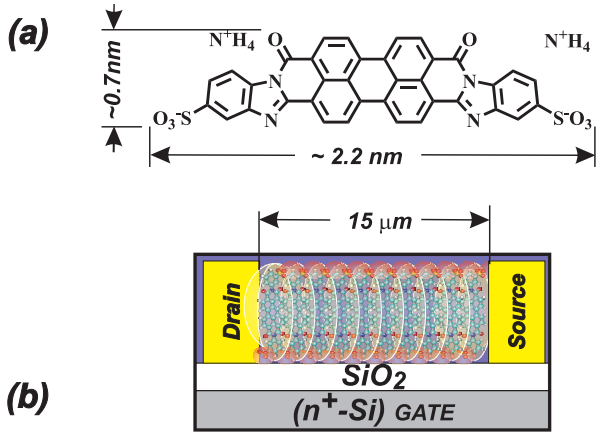}
\end{figure}
\begin{figure}[tbp]
\centering \includegraphics[width=0.48 \textwidth]{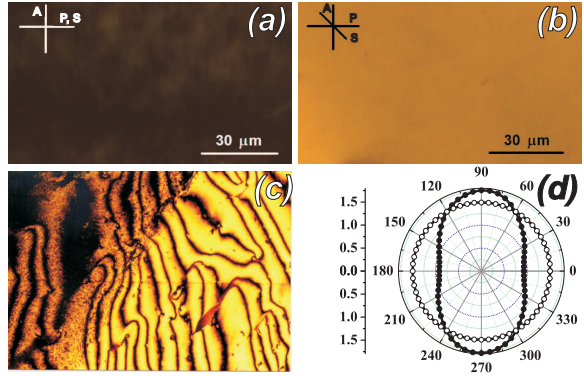}
\caption{(Color online) V20 film formed by shear deposition and viewed
between two crossed polarizers P and A with the direction of shear S\
parallel to the polarizer P (a) and under the angle of 45 degrees (b).
Polarizing microscope texture of the nematic V20 cell (7.4$\%$); right
bottom part is covered by a glass; the left upper part is allowed to dry
(c). The intensity of light transmitted by a dry V20 layer deposited from
the nematic phase (7.4\%, filled circles) and from the isotropic solution
(3.5\%, open circles) as the function of the angle between the shear
direction and polarization of light; normal incidence; wavelength 567 nm
that corresponds to the maximum of absorbance. }
\label{Textures}
\end{figure}

\begin{figure}[b]
\centering \includegraphics[width=0.48 \textwidth]{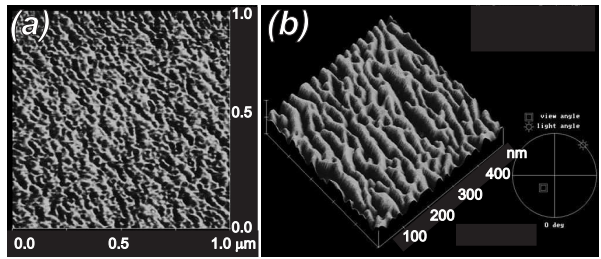}
\caption{Atomic Force Microscopy of an aligned layer of V20 deposited from
the nematic phase at the SiO$_{2}$ layer (a) and at the AP mica (b).}
\label{AFM}
\end{figure}

We used LCLC material Violet 20 (V20, Fig. \ref{Molecular_structure}a). The perylene-based core of V20 is large, to enhance the $\pi-\pi $ attractions and increase the mobility of charge carriers \cite{vandeGraats}.
The nematic phase was formed in aqueous solutions with concentration (7.4-8.0) $wt\%$ \cite{Schneider_Artyushkova}. The solution was deposited onto the studied substrates and shear-aligned using either a vertical spin coating (1500 rpm) \cite{Kobayashi} or an applicator rod. The film was dried at room temperature \cite{Schneider_Artyushkova}. The films are highly dichroic and birefringent, Fig. \ref{Textures}a,b. Film deposition from more dilute isotropic solutions produce no long-range order, Fig. \ref{Textures}d. To illustrate that the oriented structure is preserved during drying, we prepared an unaligned sample of V20 and covered one part of it with a glass plate; the remaining part was left exposed to air to facilitate drying. The two part show a continuous textural transition, Fig. \ref{Textures}c. The long-range orientational order of aggregates with the V20 planes perpendicular to the shear direction is confirmed by angular dependency of polarized light absorption, Fig. \ref{Textures}d and by atomic force
microscopy (AFM), for V20 deposited onto a glass covered with SiO$_{2}$ (that mimics the surface of the OFET device), Fig. \ref{AFM}a, and onto the commercially available AP mica (Bioforce Laboratory, Inc.), Fig. \ref{AFM}b. Although defects of packing are evident, most notably Y-shaped bifurcations, the overall percolating network of aggregates is well connected and aligned along the shear direction. In the perpendicular direction, there is no well-defined periodic structure, as the chromonic aggregates form bundles of variable width (10-100) nm. X-ray data show that the intermolecular spacing of 0.34 nm along the aggregates is the same in the aqueous solutions and in dry films.

The charge-carrier transport properties of solid films were studied in two geometries, (A) in a simple two electrodes lateral gap geometry \cite{Saric} and (B) in the three-electrode field-effect transistor geometry.

\begin{figure}[b]
\centering \includegraphics[width=0.48 \textwidth]{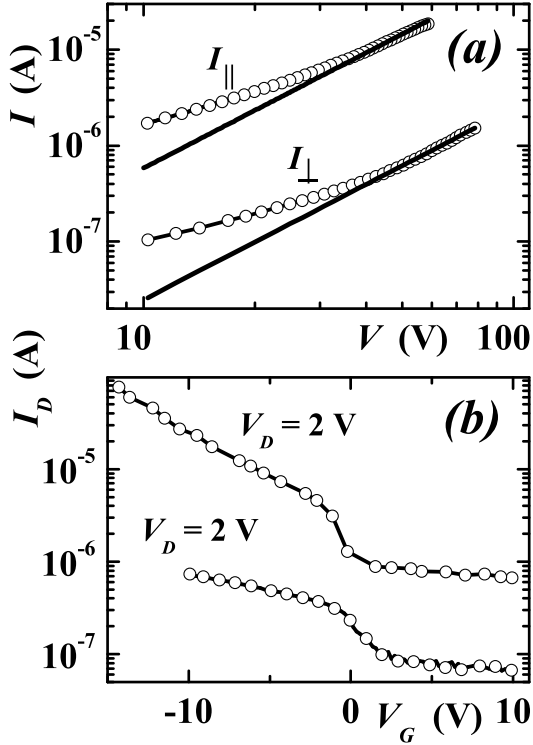}
\caption{(a) Double logarithmic plot of the current-voltage charateristics
measured in lateral diode geometry. The solid line shows quadratic power
low. (b) $I_{D}-V_{G}$ characteristics measured in two diferent samples of
LCLC-based OFET with aggregates aligned perpendicularly to the source/drain
electrodes.}
\label{Transfer_characteristics}
\end{figure}

\textbf{(A). Two electrode geometry.} The devices were fabricated by depositing LCLC film onto the gold electrodes. The two gold electrodes were vacuum evaporated and patterned by photolithography on glass substrate. The channel length and width were $L=15\mu m$ and $W=40mm$, respectively. The
measurements were performed in a vacuum box ($10^{2}$ Pa) at room temperature. The LCLCL aggregates were either parallel or perpendicular to the gap, to explore the transport anisotropy. The current $I_{\Vert }$ measured parallel to the LCLC aggregates is much higher than the current  $I_{\perp }$ in the perpendicular direction, Fig. \ref{Transfer_characteristics}a. The corresponding charge carrier mobilities
were determined from a charge-limited current (SCLC) regime in the current ($I$) -- voltage ($V$) curves as described in \cite{Saric} (Eq. (2), in which the relative dielectric permittivity of V20 was assumed to be 3, see \cite{Rybak}):  $\mu_{_{_{SCLC\Vert}}}=5\times 10^{-3}$ $cm^{2}V^{-1}s^{-1}$ and  $\mu_{_{SCLC\perp }}=3\times 10^{-5}$ $cm^{2}V^{-1}s^{-1}$.

\textbf{(B). Three electrode geometry.} The bottom-contact bottom-gate LCLC OFET devices, Fig. \ref{Molecular_structure}b, were fabricated on a highly doped silicon wafer ($\sigma =1Ohms/\square $) acting as the substrate and the gate electrode. On top of this substrate, a gate-insulating layer of silicon dioxide (SiO$_{2}$) has been thermally grown. Gold source and drain electrodes were then vacuum evaporated and patterned by photolithography. The channel length and width were $L=15\mu m$ and $W=40mm$, respectively; the capacitance of the gate dielectric was $C_{i}=15.5nF/cm^{2}$. All measurements were performed in a vacuum box ($10^{2}$ Pa) at room temperature. The drain current versus gate voltage ($I_{D}-V_{G}$) characteristics for the LCLC-based OFET measured along the aggregates show a semiconducting behavior, Fig. \ref{Transfer_characteristics}b. The drain-source current $I_{D}$ increases with the negative gate bias $V_{G}$ indicating a p-channel behavior. Many perylene diimide derivatives yield n-channel behavior\cite{Schmidt,Molinari} featuring the electron OFET mobility up to $2 cm^{2}V^{-1}s^{-1}$. However some derivatives exhibit p-channel \cite{Wang} and ambipolar channel transport \cite{Singh}. It was shown that the energy of HOMO and LUMO levels in the derivatives can vary quite considerably  with the type of substituents so that the behavior can be tuned from n- to p-type \cite{Delgado}. It should not be surprising that no n-channel behavior is observed in our OFETs since the films were prepared from a water solution and residual water can quench the electron transport. The details of charge transfer in LCLC solid films remain to be understood.

The OFET mobility $\mu _{_{FE}}$ has been determined in the linear regime of the $I_{D}-V_{G}$ characteristics (at low source-drain voltages). The applied source-gate electric field in this regime is much larger than the in-plane source-drain field, which results in an approximately uniform density of charge carriers in the conductive channel. The OFET mobility in linear regime is calculated as

\begin{equation}
\mu _{_{FE}}=\frac{L}{WC_{i}V_{D}}\frac{\partial I_{D}}{\partial V_{G}},
\label{ID}
\end{equation}

\noindent where $V_{D}$ is the source-drain voltage, $C_{i}$ is the capacitance of the gate dielectric. Using Eq. (\ref{ID}) and data from curve 1 in Fig. \ref{Transfer_characteristics}b, one calculates $\mu _{_{FE\parallel}}=3\times 10^{-2}$ $cm^{2}V^{-1}s^{-1}$ along the aggregates, which is relatively high for a device that was not optimized by any special layer at the gate-LCLC interface.  However, the measured OFET mobilities depend strongly on the details on sample fabrication, prehistory and ambient conditions. Fig. \ref{Transfer_characteristics}b shows that the $I_{D}$ - $V_{G}$ curves and the resulting mobilities might differ by several orders of magnitude if measured for two devices constructed and handled in a similar way. Moreover, when the temperature of the samples was reduced below $\sim 220$ K or the pressure was decreased to $0.1$ Pa, the electric currents vanished. All these features point towards the critical role of the residual water associated with the LCLC molecules in the films that determines the structure of aggregates, their wetting at the different substrates and thus the general performance of the devices. Optimization of LCLC based OFET would require a special control of the water component in the film.

The high charge mobility in LCLC films is naturally connected to the very essence of chromonic aggregation. The "fast conductive channels" that provide percolative passages for the charge transport, Fig. \ref{AFM}, are initially formed in water and then transferred to the substrate. The highly ordered and packed aggregates require less activation for charge hopping between the molecules. The hopping rate between the LCLC molecules is high because of their face-to-face parallel arrangement with separation that is smaller than in conventional "non-aggregated" organic semiconductors. The LCLC aggregates are intrinsically polydisperse. The charge transport is expected be dominated by longer aggregates that represent extended 1D traps for charge carriers; their escape into the less ordered regions formed by shorter aggregates is hindered by a mismatch in the corresponding HOMO/LUMO levels. This confinement feature is known to enhance the mobility \cite{Tam}. The strongly anisotropic conductivity is naturally related to the unidirectional alignment of the aggregates with a strongly reduced order in perpendicular direction, Fig. \ref{AFM}.

The LCLCs represent a distinct and promising class of materials for molecular electronics. The directional $\pi-\pi$ interactions responsible for the very existence of the orientational order in LCLC are also the driving force of electric conductivity along the aggregates. The LCLCs combine the very high packing density associated with a 0.34 nm separation within the aggregates, with the ability to form oriented structures transferable from the water solution into the dry deposited films. Note that the films made from water solutions of similar sulphonated derivatives of phthalocyanines \cite{Chaidogiannos,Nespurek} also were shown to exhibit excellent transistor properties, although the potential liquid crystalline properties
were not employed in the film preparation. In the case of V20, the liquid crystalline properties are essential in both formation and function of the organic semiconductor. The LCLCs allow one a simple manufacturing of OFETs by spin-coating or printing technology; the carrier OFET mobility is anisotropic and high, even in the non-optimized conditions, which holds a major promise for future improvements. Our current studies are focused on the role of residual water in LCLC films which appear to be a major factor controlling charge transport.

We thank A. Nych, T.\ Schneider, and A. Smith for their help in experiments. The work was supported by NSF Materials World Network on Lyotropic Chromonic Liquid Crystals DMR076290, the Science and Technology Center in Ukraine (project 5258) and National Academy of Science of Ukraine 1.4.1B/10.

\end{document}